\documentclass[aps,prd,amsmath,amssymb,footinbib]{revtex4-1}

\usepackage{graphicx}
\usepackage{natbib}

\usepackage{bm}
\usepackage{url}
\usepackage{textcase}
\usepackage{subcaption}
\usepackage{float}
\usepackage{amsthm}
\usepackage{braket}
\usepackage{color}

\begin{document}

\title{Rotating tensor-multi-scalar solitons}

\author{Lucas G. Collodel}
 \email{lucas.gardai-collodel@uni-tuebingen.de}

\affiliation{Theoretical Astrophysics, Eberhard Karls University of T\"ubingen, T\"ubingen 72076, Germany}
\affiliation{Institut f\"{u}r Physik, Universit\"{a}t Oldenburg, Postfach 2503 D-26111 Oldenburg, Germany}

\author{Daniela D. Doneva}
\email{daniela.doneva@uni-tuebingen.de}
\affiliation{Theoretical Astrophysics, Eberhard Karls University of T\"ubingen, T\"ubingen 72076, Germany}
\affiliation{INRNE - Bulgarian Academy of Sciences, 1784  Sofia, Bulgaria}

\author{Stoytcho S. Yazadjiev}
\email{yazad@phys.uni-sofia.bg}
\affiliation{Theoretical Astrophysics, Eberhard Karls University of T\"ubingen, T\"ubingen 72076, Germany}
\affiliation{Department of Theoretical Physics, Faculty of Physics, Sofia University, Sofia 1164, Bulgaria}
\affiliation{Institute of Mathematics and Informatics, 	Bulgarian Academy of Sciences, 	Acad. G. Bonchev St. 8, Sofia 1113, Bulgaria}

\begin{abstract}
In the context of a special class of tensor-multi-scalar theories of gravity for which the target-space metric admits 
an isometry under which the theory is invariant, we present rotating vacuum solutions, namely with no matter fields. These objects behave like nontopological solitons, whose primary stability is due to the conserved charge arising from the global symmetry. We consider theories with two and three scalar fields, different Gauss curvatures and quartic interaction coefficients. As it occurs for boson stars, their angular momentum is quantized. 
\end{abstract}

\maketitle

\section{INTRODUCTION}
The rapid advance in the astrophysical observations, including the currently regular detection of gravitational wave merger events, finally opens the door towards the possibility to set tight constraints on the allowed deviations from general relativity (GR) in the strong field regime. This calls for further advance in the theoretical modeling of compact objects in alternative theories of gravity in order to be able to explore the possible deviations from the Einstein's theory of gravity and to test them observationally. There is a whole plethora of GR modifications and perhaps the most widely explored is the case when we have additional scalar degree(s) of freedom \cite{Berti2015,Doneva2018} motivated from various theoretical considerations, such as higher dimensional theories, theories trying to unify all the interactions, etc. The focus of the present paper falls exactly in this area dealing with the less explored case of multiple scalar fields \cite{Damour1992,Horbatsch2015}. Such tensor-multi-scalar theories are on one hand complicated and difficult to handle, but on the other hand they  posses a lot of freedom and possibilities to construct new and very interesting solutions \cite{Yazadjiev2019,Doneva2019a,Doneva2019}. In addition, they are in agreement with the observations and do not have severe intrinsic mathematical problems.

Despite the great interest in compact objects in alternative theories of gravity, most of the solutions are obtained either in the static or in the slowly rotating regime due to the fact that the reduced field equations become very involved if we drop the spherical symmetry and turn to axially symmetric compact objects. Since practically all of the astrophysical object are rotating and some of them are expected to posses a large angular momentum close to the extremal limit for black holes or the mass-shedding limit for neutron stars, the need of further advance in the modeling of rotating compact objects in modified gravity is obvious. Even more, it was shown for example that in some alternative theories of gravity involving scalar degrees of freedom the rapid rotation can magnify the deviations from general relativity and offer the possibility for new independent tests the strong field regime of gravity \cite{Doneva2013a}.

A very interesting class of objects are the boson stars, possessing various degrees of compactness, some of which arbitrarily close to $1/2$ \cite{PhysRevD.85.024045,HARTMANN2012120}, and are serious contenders for black hole mimickers \cite{MIELKE2000185,doi:10.1142/S0218271806009637,Vincent_2016}. These objects are the realization of a complex scalar field bound by its own gravitational field. The $U(1)$ global symmetry underlying the theory gives rise to a conserved Noether current and charge \cite{PhysRev.172.1331,PhysRev.187.1767} which allows for stable solutions. They do not contain a hard surface, and the scalar field extends to infinity, although falling exponentially, permitting one to define an effective radius for the star. Therefore, and because its composing bosons only interact with matter through the gravitational channel, orbits taking place inside the effective radius of the star are possible, and severely different from geodesic motion in other spacetimes in general relativity \cite{PhysRevD.90.024068,Grould_2017,PhysRevLett.120.201103}. Because the angular momentum of these gravitational solitons is quantized \cite{SCHUNCK1998389,PhysRevD.55.6081}, it is not possible to encounter nontrivial perturbative solutions of spinning stars \cite{PhysRevD.50.7721}, but the fully nonlinear field equations must be solved. Furthermore, rotation results in a change of topology for the scalar field distribution, which becomes zero on the symmetry axis, and when spinning rapidly enough, ergoregions are formed. Boson stars have been subject of extensive investigation, its existence in the parameter space, stability and excited solutions are discussed in \cite{PhysRevD.35.3640,LEE1989477,PhysRevD.24.2111,PhysRevD.72.064002,PhysRevD.77.064025,PhysRevD.96.084066}. Remarkably, because the scalar field does not share the isometries of the spacetime, it is possible to evade the no-hair theorems when the angular velocity of a Kerr black hole matches the angular parameter $m/\omega_s$ of a marginal scalar cloud, giving rise to a hairy black hole as reported in \cite{PhysRevLett.112.221101}. An updated review on boson stars can be found in \cite{Liebling2017}.

It was recently shown that in tensor-multi-scalar theories soliton solutions exist when the target space metric admits Killing fields with a periodic flow \cite{Yazadjiev2019}. In the simplest case when this target space metric is the flat one, the field equations describing the solutions coincide with the corresponding ones for boson stars, but the phenomenology can be much richer for other non-flat target space metrics. In addition, in these theories mixed configurations of solitons and neutron stars exist \cite{Doneva2019a}. A natural next step is to generalize the solutions in \cite{Yazadjiev2019} to the case of rapid rotation that is the focus of the present paper.

In Section \ref{s1} we present the basic formalist behind the construction of solitons in tensor-multi-scalar theories. The numerical setup is briefly discussed in Section \ref{sec:NumericalSetup} and the results are presented in Section \ref{sec:Results}. The paper end with Conclusions.  The reduced field equations are given in a separate Appendix.
 
\section{THEORY AND CONSERVED CHARGES}
\label{s1}

In the TMST of gravity, the gravitational interaction is mediated not only by the metric of the spacetime but also by $N$ additional dynamical scalar fields $\varphi^a$. The scalar fields $\varphi^a$ can be considered as a generalized coordinates on  a coordinate patch of
an abstract target manifold ${\cal E}_N$ supplemented  with a positive definite metric $\gamma_{ab}(\varphi)$. The general vacuum action of the TMST presented in the Einstein frame is given by

\begin{equation}
\label{action}
S=\frac{1}{16\pi G_{\ast}}\int \left[R-2g^{\mu\nu}\gamma_{ab}(\varphi)\partial_\mu\varphi^a\partial_\nu\varphi^b-4V(\varphi)\right]\sqrt{-g}d^4x. 
\end{equation}
where $R$ is the Ricci curvature with respect to the Einstein frame metric $g_{\mu\nu}$ and $g$ is its determinant. The Einstein frame metric is connected to the physical Jordan frame one $\tilde{g}_{\mu\nu}$ via a conformal transformation $\tilde{g}_{\mu\nu}=A^2(\varphi) g_{\mu\nu}$.

Varying the action with respect to the spacetime metric $g^{\mu\nu}$ we find the Einstein frame 
field equations 

\begin{equation}
\label{efe}
R_{\mu\nu}=2\gamma_{ab}\partial_\mu\varphi^a\partial_\nu\varphi^b+2V(\varphi)g_{\mu\nu},
\end{equation}
while varying the system with respect to the field $\varphi^a$ results in the corresponding equations for the scalar fields,
\begin{equation}
\label{kge}
\Box\varphi^a=-\gamma^a_{bc}(\varphi)g^{\mu\nu}\partial_\mu\varphi^b\partial_\nu\varphi^c+\gamma^{ab}(\varphi)\frac{\partial V(\varphi)}{\partial\varphi^b},
\end{equation}
where $\gamma^a_{bc}(\varphi)$ is the Christoffel symbols for the metric $\gamma_{ab}(\varphi)$ of the target space. If this metric admits a Killing field $K^a$ with a periodic flow, then the scalar field's kinetic term in the theory (\ref{action}) is invariant under its action. Similarly, if the field's potential and the conformal factor are also invariant under the flow, i.e. $\mathcal{L}_KV(\varphi)=K^a\partial_aV(\varphi)=0$ and $\mathcal{L}_KA(\varphi)=K^a\partial_aA(\varphi)=0$, then $K^a$ is the generator of a one-parameter group of point transformations for which the whole theory is invariant, and a corresponding Jordan frame Noether current therefore exists, given by \cite{Yazadjiev2019}
\begin{equation}
\label{nc}
\tilde{j}^\mu=\frac{1}{4\pi G(\varphi)}\tilde{g}^{\mu\nu}K_a\partial_\nu\varphi^a,
\end{equation} 
with $G(\varphi)=G_{*}A^2(\varphi)$.

In the present paper we are interested in stationary  and axisymmetric solutions, thus we adopt the quasi-isotropic Lewis-Papapetrou metric in adapted spherical coordinates $(t,r,\theta,\varphi)$, for which the line element reads
\begin{equation}
\label{metric}
ds^2=-fdt^2+\frac{l}{f}\left[g\left(dr^2+r^2d\theta^2\right)+r^2\sin^2\theta\left(d\phi-\frac{\omega}{r}dt\right)^2\right].
\end{equation}
The metric coefficients are all functions of $r$ and $\theta$ alone, and this spacetime then possesses two Killing vector fields, namely $\xi=\partial_t$, and $\eta=\partial_\phi$. The reduced field equations obtained after imposing these symmetries are quite lengthy and are given in a separate Appendix \ref{section:Appendix}

\subsection{N=2 model}

A two dimensional manifold can always be brought to a conformally flat form, $\gamma_{ab}=\Omega^2_2(\varphi)\delta_{ab}$, where $\delta_{ab}$ is the Kroenecker delta. In these coordinates, the Killing field with periodic flow is given by $K=\varphi^{(2)}\frac{\partial}{\partial\varphi^{(1)}}-\varphi^{(1)}\frac{\partial}{\partial\varphi^{(2)}}$. This imposes the requirement that $\Omega^2_2(\varphi)$, and similarly $U(\varphi)$ and $A(\varphi)$, depends on $\varphi^a$ through the combination $\psi^2=\delta_{ab}\varphi^a\varphi^b$.

In what follows, we are focused on maximally symmetric 2-manifolds. Henceforth, the conformal factor assumes the form
\begin{equation}
\Omega^2_2=\frac{1}{\left(1+\frac{\kappa}{4}\delta_{ab}\varphi^a\varphi^b\right)^2},
\end{equation}
where $\kappa$ is the Gauss curvature of the target manifold, whose geometry is spherical for $\kappa>0$, flat for $\kappa=0$ and hyperbolic for $\kappa<0$.

The self-interaction potential we consider is the quartic
\begin{equation}
V(\psi)=\frac{1}{2}\mu^2\psi^2+\frac{1}{4}\lambda_{(4)}\psi^4.
\end{equation}
\subsection{N=3 model}

In this model we assume a target space which is a warped product of the real line with the 2-manifold described above. The internal line element reads,
\begin{equation}
\label{s2n3}
\gamma_{ab}d\varphi^ad\varphi^b=d\chi^2+\Omega^2_3(\psi,\chi)\delta_{AB}d\varphi^Ad\varphi^B,
\end{equation}
where $\varphi^a=(\varphi^A,\chi)$ and $\Omega^2_3(\psi,\chi)=F(\chi)\Omega^2_2(\psi)$ and throughout this work we choose $F(\chi)=e^{2\chi}$.

In this case for simplicity we consider only noninteracting massive fields,
\begin{equation}
V(\psi,\chi)=\frac{1}{2}\mu^2(\psi^2+\chi^2).
\end{equation}
\subsection{Ansatz}

Solitonic solutions of this system only exist for time dependent fields $\varphi^A$, such that $\mathcal{L}_\eta\varphi^A\neq 0$. The only way these fields could depend on time without jeopardizing the stationarity of the spacetime is by using the functions defined by the Lie group transformation rules that give rise to the conserved current (\ref{nc}), promoting the one parameter to a linear function of $t$, such that 
\begin{equation}
\label{ak1}
\mathcal{L}_\eta\varphi^A=-\omega_sK^A,
\end{equation}
where $\omega_s$ is a real constant. In a similar fashion, in order to source the dragging of the spacetime, there must be a nontrivial $\phi$ dependence on the fields $\varphi^A$, for which
\begin{equation}
\label{ak1}
\mathcal{L}_\xi\varphi^A=-mK^A.
\end{equation}

These descriptions imply that 
\begin{equation}
\label{ans}
\left\{\varphi^1,\varphi^2\right\}=\left\{\psi(r,\theta)\cos(\omega_st+m\phi),\psi(r,\theta)\sin(\omega_st+m\phi)\right\},
\end{equation}
bringing the $SO(2)$ Lie group underlying the Noether symmetry to the $U(1)$ as this requires the amplitudes of both fields to be given by
the same function $\psi(r,\theta)$. For the $N=3$ case, the extra field shares the isometries of the spacetime, and $\chi=\chi(r,\theta)$. Moreover, we note that because the axial coordinate $\phi$ is compact, $m$ must be an integer so to preserve the identification of $\phi=0=\phi=2\pi$.

\subsection{GLOBAL CHARGES}

The Killing vectors associated to the metric $g_{\mu\nu}$ are themselves connected to two conserved global charges, namely mass (stationarity) and angular momentum (axisymmetry). These quantities can be calculated by means of the Komar integrals,
\begin{equation}
M=2\int_\Sigma R_{\mu\nu}n^\mu\xi^\nu dV,\qquad J=-\int_\Sigma R_{\mu\nu}n^\mu\eta^\nu dV,
\end{equation} 
bounded at spatial infinity. Here, $\Sigma$ is a spacelike asymptotically flat hypersurface and $n^\mu$ is a vector normal to $\Sigma$ such that $n^\mu n_\mu=-1$. The metric (\ref{metric}) implies that $n^{\mu}=(\xi^\mu+\omega/r\eta^\mu)/\sqrt{f}$. Substituting the Ricci tensor with aid of eq. (\ref{efe}) and (\ref{ans}), and noticing that the volume element is $dV=\sqrt{-g/f}drd\theta d\phi$ we arrive at
\begin{equation}
\label{intmj}
M=\int(2\mathfrak{j}-U)\sqrt{-g}drd\theta d\phi, \qquad J=\int \mathfrak{j}\sqrt{-g}drd\theta d\phi,
\end{equation}
where we defined the angular momentum density
\begin{equation}
\mathfrak{j}=2m\frac{\left(m\omega+\omega_sr\right)\Omega^2_N\psi^2}{fr}.
\end{equation}
The subscript takes values $N=\{2,3\}$ depending on the model one considers. 

The conserved current \eqref{nc} leads to a conserved charge which is given by
\begin{equation}
\label{nnc}
Q = \int_\Sigma  \tilde{j}_\mu n^\mu dV, 
\end{equation}
and by applying the same substitutions we did for the Komar integrals, we arrive at
\begin{equation}
\label{nnc2}
Q = \int \frac{\mathfrak{j}}{m}\sqrt{-g}drd\theta d\phi,
\end{equation}
meaning that, just like for boson stars, the rotating soliton-like vaccum of these TMST theories features a quantized angular momentum $J=mQ$.

The mass and angular momentum, being global charges, can also be measured at spatial infinity and hence be extracted asymptotically from the metric, emphasizing the equivalence between the Komar and ADM charges in asymptotically flat spacetimes
\begin{equation}
\label{asympmj}
M=\frac{1}{2}\lim_{r\rightarrow\infty}r^2\partial_rf, \qquad J=\frac{1}{2}\lim_{r\rightarrow\infty}r^2\omega.
\end{equation}

\subsection{Boundary Conditions}
In order to find solitonic solutions in this theory we need to establish the correct boundary conditions which guarantee that the spacetime is regular and asymptotically Minkowsky, and the scalar fields' solution is localized. At the origin, regularity requires that
\begin{align}
\partial_rf\rvert_{r=0}=0,\qquad \partial_rl\rvert_{r=0}=0,\qquad g\rvert_{r=0}=1,\qquad \omega\rvert_{r=0}=0, \qquad \psi\rvert_{r=0}=0, \qquad \partial_r\chi\rvert{r=0}=0,
\end{align}
and an expansion of $\psi$ around $r\sim 0$ yields
\begin{align} 
\psi(r,\theta)&=\psi_0r^m\sin^m\theta+\mathcal{O}(r^{m+1}),
\end{align}
where we use the leading term to define the central value $\psi_0\equiv\frac{1}{m!}\frac{\partial^m\psi(0)}{\partial r^m}$, which is useful to parametrize the solutions.

The spacetime is asymptotically flat, and the fields $\varphi^a$ vanish at infinity,
\begin{align}
f\rvert_{r\rightarrow\infty}=1,\qquad l\rvert_{r\rightarrow\infty}=1,\qquad g\rvert_{r\rightarrow\infty}=1,\qquad \omega\rvert_{r\rightarrow\infty}=0, \qquad \psi\rvert_{r\rightarrow\infty}=0, \qquad \chi\rvert_{r\rightarrow\infty}=0.
\end{align}
Solving for $\psi$ near infinity gives
\begin{align}
\psi(r\sim\infty)\propto\frac{1}{r}\exp{\left(-\sqrt{\mu^2-\omega^2}r\right)},
\end{align}
so that bound solutions can only occur if $\omega_s^2\leq\mu^2$, and the system becomes trivial when the equality holds.

On the symmetry axis, the elementary flatness conditions sets $g\rvert_{\theta=0}=1$. The other fields are, once again, determined as to guarantee regularity,\textbf{}
\begin{align}
\partial_\theta f\rvert_{\theta=0}=0,\qquad \partial_\theta l\rvert_{\theta=0}=0,\qquad g\rvert_{\theta=0}=1,\qquad \partial_\theta\omega\rvert_{\theta=0}=0, \qquad \psi\rvert_{\theta=0}=0, \qquad \partial_\theta\chi\rvert_{\theta=0}=0,
\end{align}
and finally, because we are describing a system with even parity, all angular derivatives must vanish on the equatorial plane,
\begin{align}
&\partial_\theta f\rvert_{\theta=\pi/2}=0,\qquad \partial_\theta l\rvert_{\theta=\pi/2}=0,\qquad \partial_\theta g\rvert_{\theta=\pi/2}=0,\qquad \partial_\theta\omega\rvert_{\theta=\pi/2}=0,\qquad \partial_\theta\psi\rvert_{\theta=\pi/2}=0,\qquad \partial_\theta\chi\rvert_{\theta=\pi/2}=0.
\end{align}

\section{Numerical Setup}\label{sec:NumericalSetup}

The set of PDE, displayed in Appendix \ref{section:Appendix}, is solved with aid of the program package FIDISOL \cite{SCHONAUER1990279}, which employs a finite-difference method over a user provided grid upon which the equations are discretized. A Newton scheme is used for linearizing the resulting equations, and the final linear system is solved iteratively. We note that the transformations, $r=\bar{r}/\mu$,  $\omega_s=\bar{\omega}_s\mu$ and $\lambda_{(4)}=\lambda\mu^2$ leave the equations independent of the mass $\mu$ and we use it for normalization of the quantities displayed in the figures. Since some boundary conditions are set at spatial infinity, we compactify the radial coordinate via the mapping
\begin{align}
\tilde{r}=\frac{\bar{r}}{\bar{r}+1},
\end{align}
and construct nonuniform rectangular grids with the Cartesian product $[0,1]\times[0,\pi/2]$, containing $125\times 50$ points. The maximum error we allow for each field is of order $10^{-7}$.

The first solution is obtained near the critical value $\omega_s^2/\mu^2\sim 1$, where spacetime is nearly flat and the fields' amplitudes are infinitesimal, and therefore $\psi_0\ll 1$. This result is then used as an initial guess for the next solution, where the input parameter $\omega_s$ is decreased slightly, and this process continues until there is no more convergence. Thereon we proceed by doing a polynomial interpolation of three previous solutions to extrapolate an initial guess for a higher value of $\psi_0$. The approach is precisely the same for the $N=3$ theory, since there is no input parameter for the $\chi$ field, as its existence is supported by the primary fields already present in the $N=2$ case.

\section{Results} \label{sec:Results}

\subsection{N=2 model}

\begin{figure}[h!]
\includegraphics[scale=1.5]{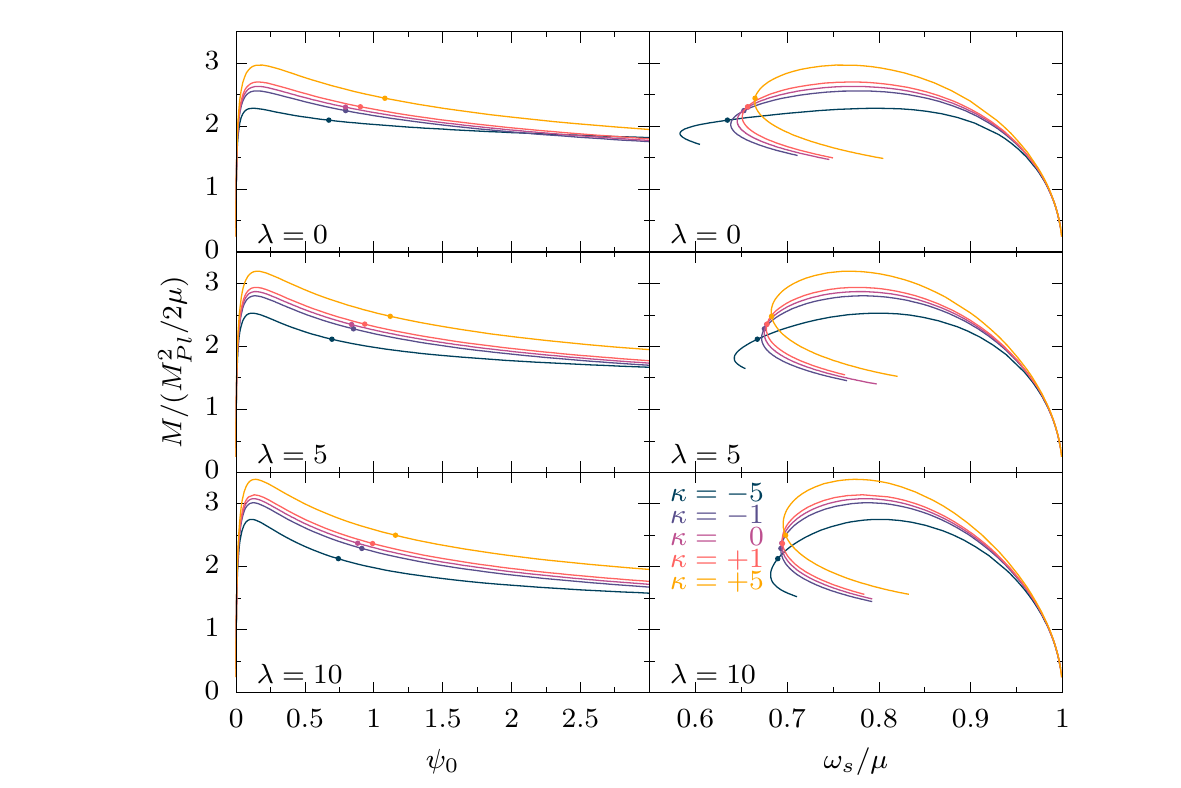}
\caption{\emph{Left Panel:} Normalized mass of the system in the $N=2$ theory against the central value of the scalar field $\psi_0$. \emph{Right Panel:}   
 Normalized mass of the system in the $N=2$ theory against the field's normalized frequency $\omega_s/\mu$.  }
\label{Fn2m1M}
\end{figure}

Several families of solutions are constructed by varying the parameters that define the theories, namely $\kappa$ and $\lambda$. In Fig. \ref{Fn2m1M} we present the normalized mass $M/\frac{M_{Pl}}{2\mu}$, where $M_{Pl}$ is the Planck mass, against the central value of the scalar field $\psi_0$ (left panel) and against $\omega_s/\mu$ (right panel), for different values of $\kappa$ and $\lambda$ but for a fixed winding number, $m=1$. As either the Gauss curvature or the quartic interaction coefficient takes larger values, the maximum of the mass for the particular family of solutions increases, as well as the lower bound for the field's frequency $\min(\omega_s)$. The smaller the value of $\kappa$ is, the more sensitive to $\lambda$ it becomes. Furthermore, we observe on the top left panel that the two curves corresponding to the smallest values of curvature intercept the subsequent two for a certain central value, displaying thenceforth higher masses for fixed $\psi_0$. Such behavior is attenuated as $\lambda$ takes higher values as the negative slope of the curves after the point of maximum mass decreases, and one needs to access the solutions at much higher central values to verify the intersections. The filled circles indicate the onset of solutions containing ergoregions, which always occur beyond the solution of maximum mass and, more importantly, maximum charge in a branch already thought to be unstable. Furthermore, as the Gauss curvature increases, the first ergoregion solution tends to occur for higher values of $\psi_0$, with an exceptional region in a small neighborhood around $\kappa=0$.

\begin{figure}
\includegraphics[scale=1.5]{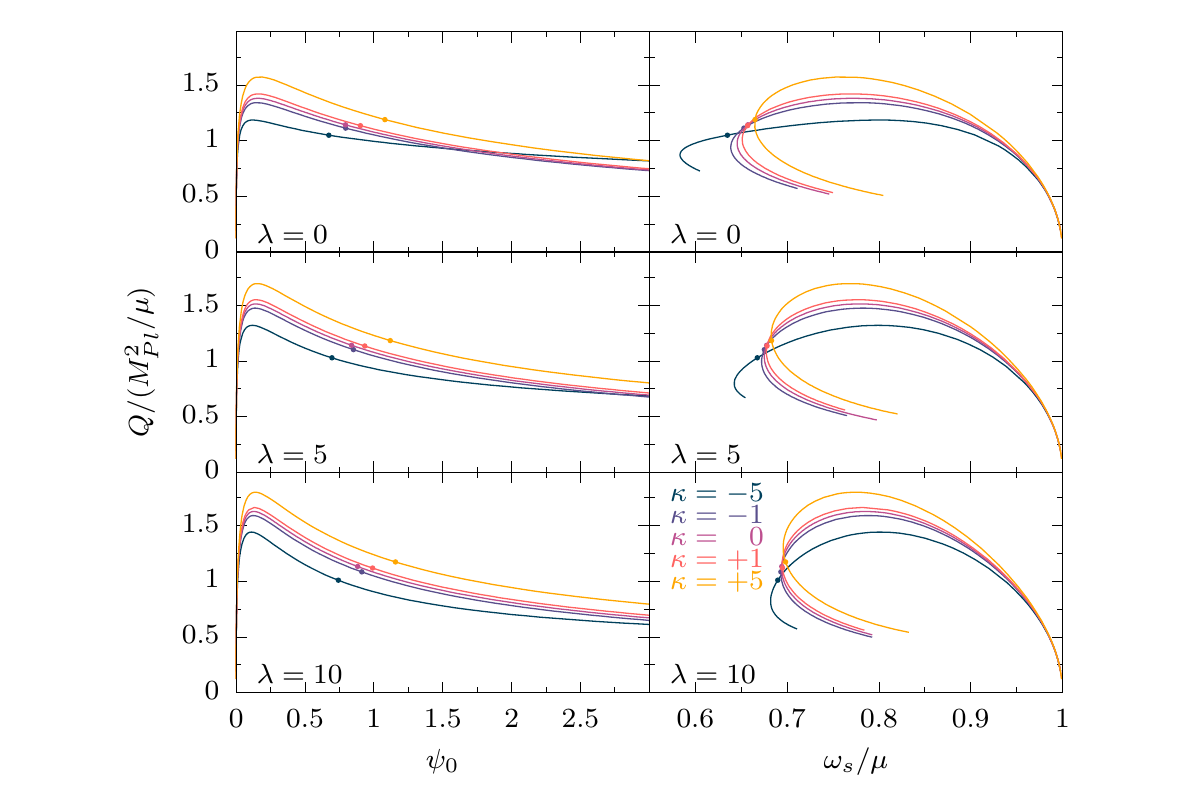}
\caption{\emph{Left Panel:} Normalized charge of the system in the $N=2$ theory against the central value $\psi_0$. \emph{Right Panel:}   
 Normalized charge of the system in the $N=2$ theory against the field's normalized frequency $\omega_s/\mu$.  }
\label{Fn2m1Q}
\end{figure}

\begin{figure}
\includegraphics{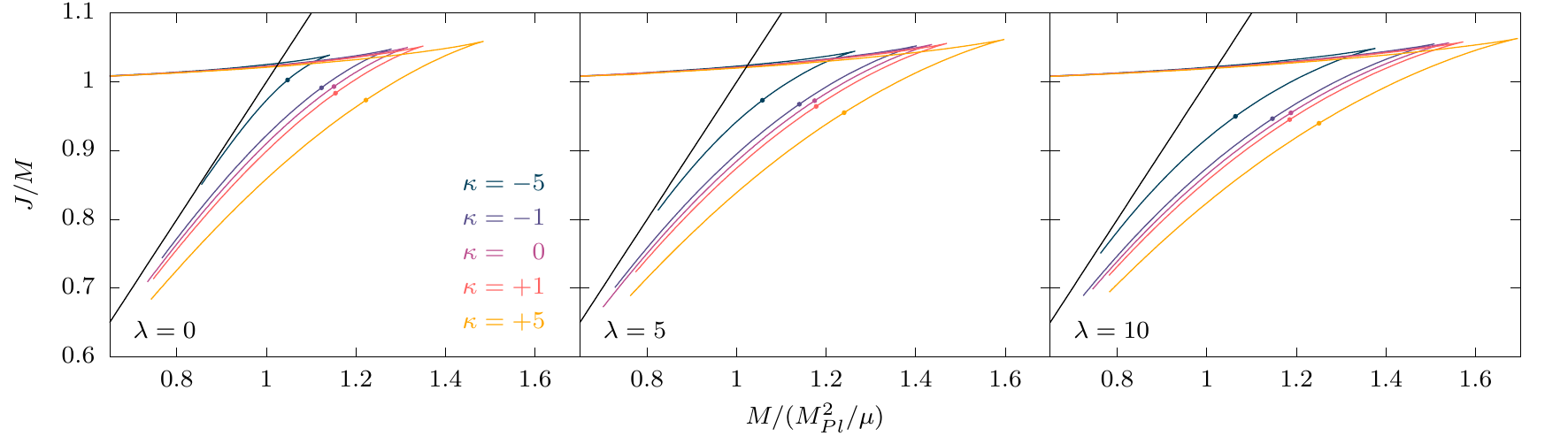}
\caption{ The Kerr parameter $a=J/M$ against the normalized mass for solutions with different Gauss curvature and quartic interaction. The ratio between the vertical and horizontal axis gives the spin parameter, which equals one for the black line.  }
\label{Fn2m1axM}
\end{figure}

The conserved normalized Noether charge, $Q/\frac{M_{Pl}}{\mu}$, of these solutions is displayed in Fig. \ref{Fn2m1Q}. The qualitative behavior drawn here is much the same as for the mass, depicted above, but slightly more accentuated. The charge drops more rapidly than the mass after its maximum as $\psi_0$ increases, as well as the curves intersect themselves for smaller $\psi_0$. The angular momentum in these systems is numerically identical to the total charge, for $m=1$. The smaller $\kappa$ is, the smaller is the charge (angular momentum) for small central values, but also the mass, as noted before, and ergoregions appear earlier (when parametrized with $\psi_0$). Therefore, such spacetimes feature a stronger dragging than those of larger Gauss curvature for the same total mass. In Fig. \ref{Fn2m1axM} we show this effect by confronting the Kerr parameter $a=J/M$ with the total mass. The black line is not part of the sets of solutions but serves as reference, as its slope (spin parameter $a_*=J/M^2$) is one everywhere, and it appears to be an asymptote to all curves. In all cases, low mass solutions (small $\psi_0$ and large $\omega_s$) have a spin parameter larger than one. There is a cusp at the maximum of the mass, indicating a change in the stability of the systems, and thereafter the smaller $\kappa$ is, the faster the curve approaches the unit slope black line. 

\subsection{N=3 model}

The target spacetime in this theory admits precisely the same Killing field as before, thus the extra field $\chi$ is a function only of the radial and polar coordinates, which makes it rather special conceptually as its stability must be supported by the primary field $\psi$ because of its staticity. Moreover, it does not source any dragging of the spacetime but counteracts it inertially. Hence, for each $\kappa$, the maximum of the mass and charge achieved by the set of equilibrium solutions are smaller than in the comparable set for $N=2.$

\begin{figure}
\includegraphics[scale=1.5]{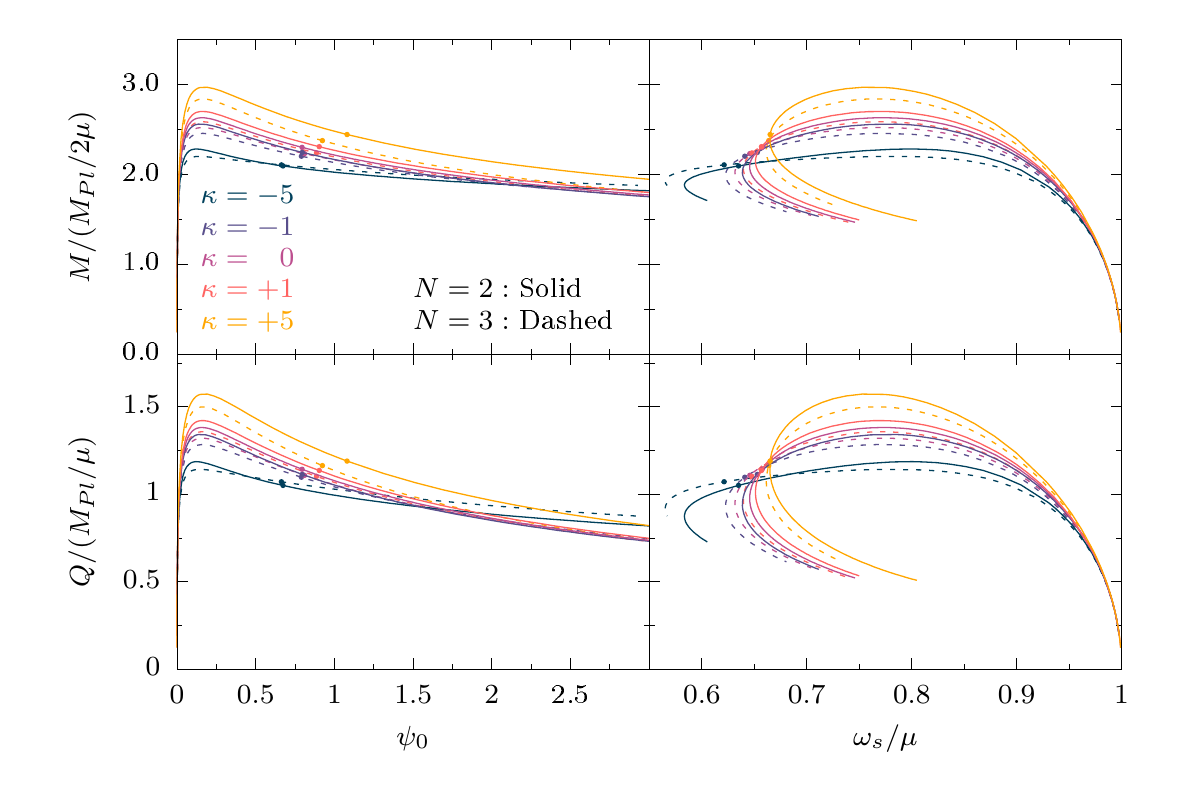}
\caption{ Mass (upper panel) and charge (lower panel) against the central value (left panel) and natural frequency (right panel) for both $N=2$ and $N=3$ theories.  }
\label{Fn2A3m1}
\end{figure}

The profile of the total mass and charge with respect to the central value of the scalar field $\psi_0$ and frequency in this theory is drawn in Fig. \ref{Fn2A3m1}, together with the $N=2$ model for comparison. Again, the full circles indicate the onset of solution with ergoregions, which take place for slightly smaller $\psi_0$ in the $N=3$ model. We note that as the central value increases, the curves intercept themselves and eventually the three fields' theory produces solutions of higher mass and charge than the two fields'. The interval in $\omega_s$ for which solutions exist is also wider in this theory.

\begin{figure}
\includegraphics[scale=.8]{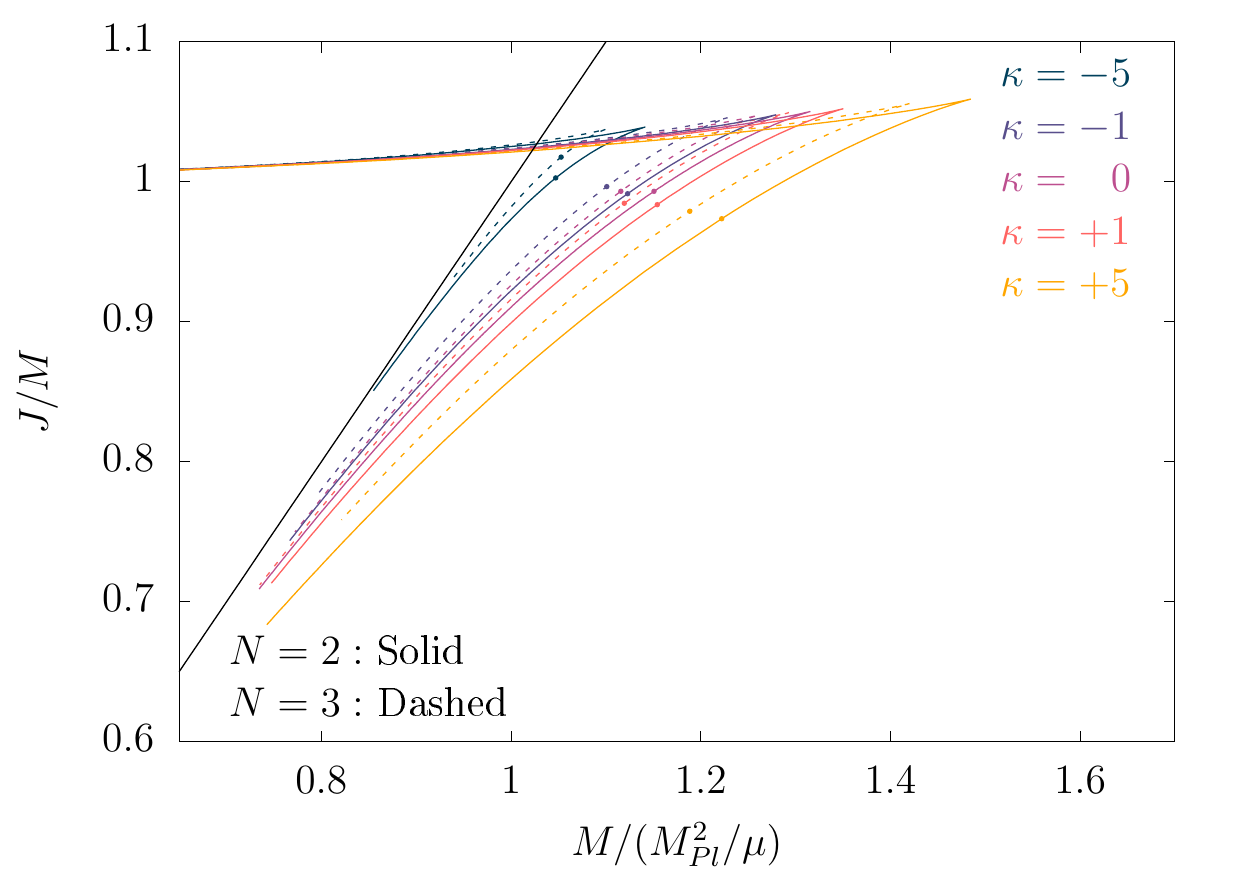}
\caption{ The Kerr parameter $a=J/M$ against the normalized mass for solutions with different Gauss curvature for both $N=3$ and $N=2$ theories with $\lambda=0$.}
\label{Fn2A3m1axM}
\end{figure}

In Fig. \ref{Fn2A3m1axM} we present the Kerr parameter against the mass, comparing different Gauss curvature sets of solutions in the $N=3$ theory with those of $N=2$ (for $\lambda=0$). Again, similar to what happens by decreasing $\kappa$, the solutions belonging to the $N=3$ theory reach a lower maximum of the mass and approach faster the asymptote.

\begin{figure}[h!]
\includegraphics[scale=1.5]{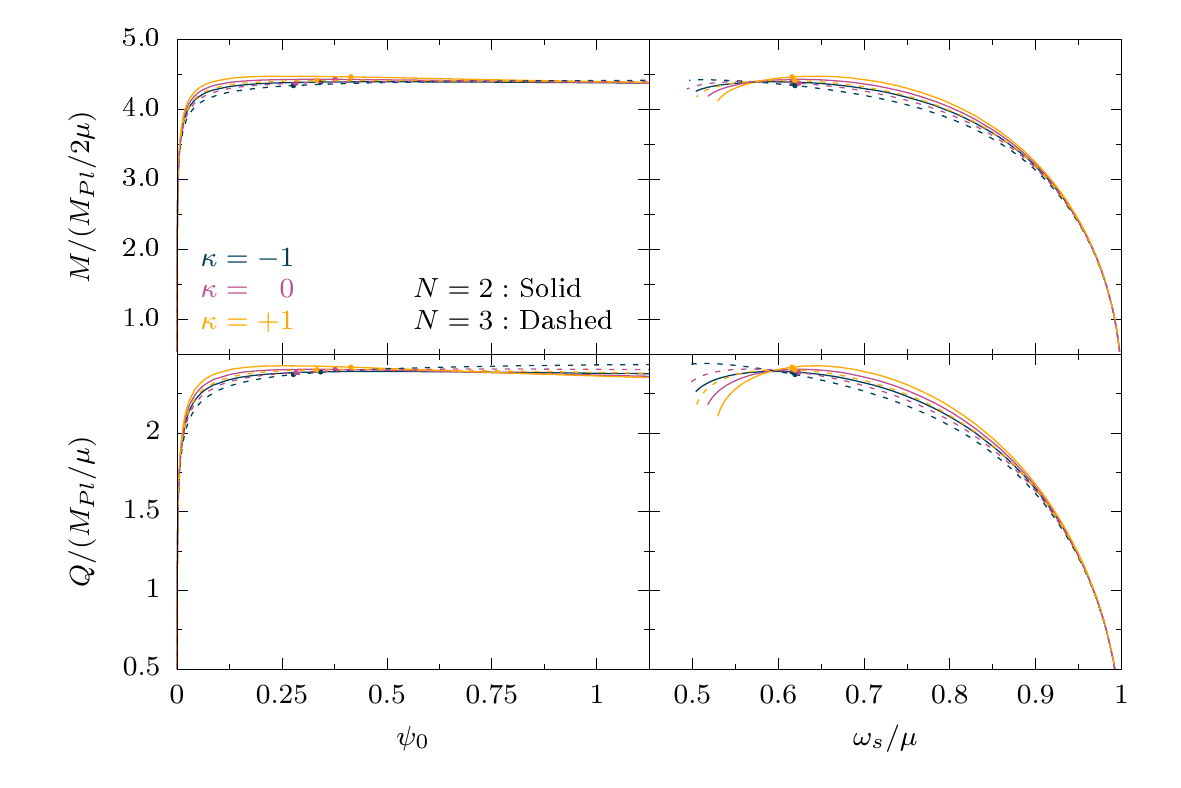}
\caption{ Mass (upper panel) and charge (lower panel) against the central value (left panel) and natural frequency (right panel) for both $N=2$ and $N=3$ theories, with winding number $m=2$.} 
\label{Fn2A3m2}
\end{figure}

The winding number $m$ is an integer input parameter, and although the angular momentum varies continuously as the central value of the scalar field increases, it takes always multiple values of the conserved Noether charge. Therefore, there is an infinite number of solution sets, each for a different integer $m$. Thus far we have focused only on the $m=1$ case, since it is the first rotational excitation the system may acquire, but for completeness we show in Fig. \ref{Fn2A3m2} some families for both theories we presented where $m=2$, displaying the mass and charge of the systems against the central value and the field's frequency.

\begin{figure}[h!]
\includegraphics[scale=.8]{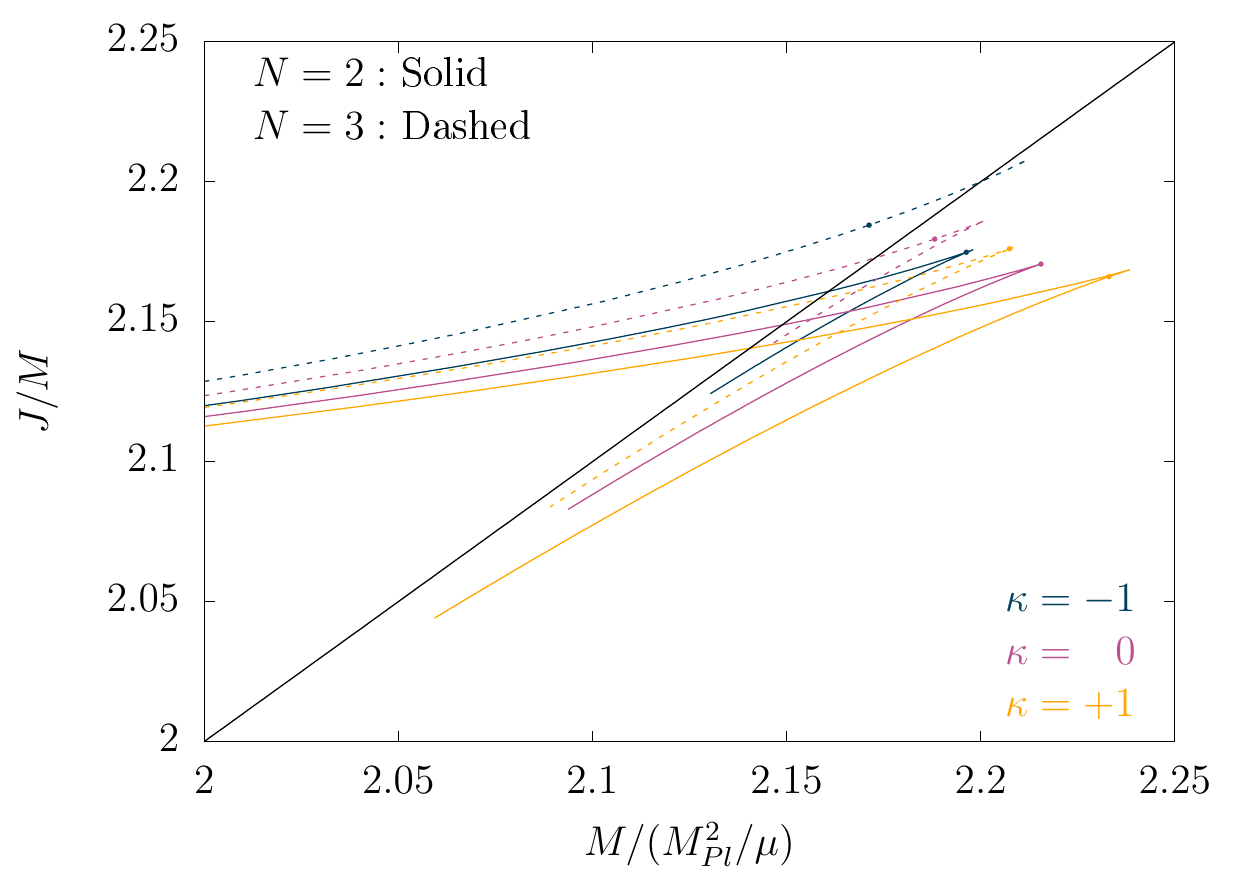}
\caption{The Kerr parameter $a=J/M$ against the normalized mass for solutions with different Gauss curvature for both $N=3$ and $N=2$ theories, with winding number $m=2$. }
\label{Fn2A3m2axM}
\end{figure}

The general behavior we observed previously for different Gauss curvatures and number of scalar fields remains, in a much more subtle fashion. The curves fall much closer to one another, and the global charges fall much slower with $\psi_0$, making it very challenging numerically to obtain solutions within the spiraling region when plotted against $\omega_s$. Due to stronger dragging effects, the equilibrium solutions are found with considerable higher masses and charges than those of $m=1$, and because the angular momentum scales with the winding number, its gain is much more prominent. In most solution sets, the onset of ergoregions happens before the solution of maximum charge, an effect that is usually seen in rotating boson stars with solitonic potential even when $m=1$ \cite{PhysRevD.77.064025,PhysRevD.85.024045}. This could trigger a different kind of instability in an otherwise stable branch \cite{PhysRevD.77.124044}. In this sense, one could expect more stable configurations in models where the Gauss curvature is higher, i.e. the bigger maximum mass stretches the domain of the stable branch and ergoregions arise only in the already unstable one. In Fig. \ref{Fn2A3m2axM} we present the Kerr parameter, which follows the same pattern as before, i.e. for small $\psi_0$ all solutions are well above the unit slope ($a_*=1$), then reach a turning point where the mass is maximum, and approach the asymptote. Interestingly, for $\kappa=-1$ and $N=3$, not only does the first ergoregion solution appear before the point of maximum charge, but it occurs for a spin parameter large than one. High spin parameters, although fairly common for planets and nuclear stars, are quite rare to achieve with compact objects, for instance rapidly rotating neutron stars are usually bound by $a_*\leq 0.7$, while the more exotic quark stars could go beyond the unit slope \cite{Lo_2011}.

\section{Conclusions}\label{sec:Conclusions}

In the present paper we have generalized the solitonic solutions in tensor-multi-scalar theories obtained in \cite{Yazadjiev2019} to the case of rapid rotating. The key essence in constructing such solutions is that we require the target space metric to admit Killing fields with a pedioc flow. Even though in the simplest case when the target space metric is the flat one these solutions coincide with the pure boson stars, strong differences can appear for non-flat target space metrics and for number of the scalar fields greater than two.  

In in the standard boson star case, the angular momentum is quantized being an integer number times the Noether charge. Naturally, this prevents from using approximate treatment of the problem, such as the slow rotation approximation, and calls for a solution of the fully nonlinear reduced field equations. We have presented solutions in the case of two ($N=2$) and three ($N=3$) scalar fields and  for winding numbers $m=1$ and $m=2$. Since the presented results are normalized to the mass of the scalar field, the input parameters are the Gauss curvature of the target space $\kappa$ and the quartic interaction coefficient $\lambda$ entering in the scalar field potential, where the pure boson star case corresponds to $\kappa=0$. The results for different values of $\kappa$ and $\lambda$, and for $N=2$ and $N=3$, are qualitatively similar. For solutions before and a bit after the maximum of the mass, the increase of the Gauss curvature is marked by an increases of the  mass and the Noether charge of the solitonic solutions while for masses much larger than the maximum one, this behavior changes and the solutions with smaller $\kappa$ have larger masses. If one studies the behavior of the Kerr parameter $a=J/M$ as a function of the soliton mass, a clear cusp is observed at the maximum mass point for all of the soliton branches which signals a change of stability.

We have examined as well the formation of an ergoregion. Such ergoregion appears for  larger mass and larger Noether charges (and thus angular momentum), and in most cases this happens for smaller central values of the scalar field when the Gauss curvature of the target space metric is reduced. For the case with two scalar fields ($N=2$) and for the examined values of the parameters, ergoregion appears for solitons belonging to the unstable part of the branch after the maximum of the mass. Ergoregion can form, though, even before the maximum mass in the case of three scalar fields ($N=3$) and for negative $\kappa$. This can potentially destabilizes the soliton solutions even before the maximum mass is reached. 

Let us comment on the possible astrophysical implications of the obtained solutions. This richness of these soliton solutions is held in the choice of the target space metric. Another factor that has to be taken into account is the presence of conformal factor that relates the quantities in the Einstein frame and the physical Jordan frame. This conformal factor enters explicitly for example in the calculation of the effective radius of the soliton \cite{Yazadjiev2019} thus different choices of $A(\varphi)$ would lead to different compactness of the object. Therefore, different observational phenomena such as mergers of solitons, electromagnetic emission connected to accretion discs, as well as the modeling of dark matter as constituent of condensed gravitational scalars, will be strongly affected both by the choice of target space metric and of $A(\varphi)$. Considering rotating soliton solutions is very important since it is an inevitable part of some of the above mentioned astrophysical models. At this stage, though, it is difficult to give a clear prescription how to distinguish between boson stars and solitons in tensor-multi-scalar theories since in the former case we can also have a variety of solutions controlled mainly by the choice of the potential. Thus further studies and modeling of the above mentioned phenomena in the framework of tensor-multi-scalar theories is important. 
 
 \section*{Acknowledgements}
LC and DD acknowledge financial support via an Emmy Noether Research Group funded by the German Research Foundation (DFG) under grant no. DO 1771/1-1. DD is indebted to the Baden-Wurttemberg Stiftung for the financial support of this research project by the Eliteprogramme for Postdocs.  SY would like to thank the University of Tuebingen for the financial support.  SY acknowledges financial support by the Bulgarian NSF Grant KP-06-H28/7. Networking support by the COST Actions  CA16104 and CA16214 is also gratefully acknowledged.

\appendix

\section{Field Equations}\label{section:Appendix}
\label{fea}
The Einstein field equations in diagonal form as we used are displayed below together with the two Klein-Gordon equations. A new parameter was introduced to keep the equations general, namely 
\begin{align}
\label{deltap}
\delta=
\begin{cases}
0 & \quad \text{if } N=2 \\
1 & \quad \text{if } N=3 \\
\end{cases}
\end{align}

\begin{align}
\label{efe1}
r^2\partial^2_rf+\partial^2_\theta f &= -\frac{1}{2}\frac{1}{fl}\Biggl[8fl^2gr^2V - 8l^2g\Omega^2_N\psi^2\left(\omega_sr+m\omega\right)^2 -2l\left(\left(r\partial_rf\right)^2+\left(\partial_\theta f\right)^2\right) -2l^2\sin^2\theta\left(\left(r\partial_r\omega\right)^2+\left(\partial_\theta\omega\right)^2\right) \nonumber \\
&+f\left(r^2\partial_rf\partial_rl+\partial_\theta f\partial_\theta l\right)+2fl\left(2r\partial_rf+\cot\theta\partial_\theta f\right)
-2l^2\omega\sin^2\theta\left(\omega-2r\partial_r\omega\right)\Biggr].
\end{align}

\begin{align}
\label{efe2}
r^2\partial^2_rl+\partial^2_\theta l &= -\frac{1}{2}\frac{1}{f^2l}\Biggl[16fl^2gr^2V-8l^2g\Omega^2_N\psi^2\left((\omega_sr+m\omega)^2l-m^2f^2\csc^2\theta\right) \nonumber \\
&-f^2\left(\left(r\partial_rl\right)^2+\left(\partial_\theta l\right)^2\right)+2f^2l\left(3r\partial_rl+2\cot\theta\partial_\theta l\right)
\Biggr].
\end{align}

\begin{align}
\label{efe3}
r^2\partial^2_rg+\partial^2_\theta g &= \frac{1}{2}\frac{1}{f^2l^2g}\Biggl[8fl^3g^3r^2V-4l^2g^3\Omega^2_N\psi^2\left((\omega_sr+m\omega)^2l-3m^2f^2\csc^2\theta\right) -l^2g^2\left(\left(r\partial_rf\right)^2+\left(\partial_\theta f\right)^2\right) \nonumber \\
&+f^2g^2\left(\left(r\partial_rl\right)^2+\left(\partial_\theta l\right)^2\right)+ 2f^2l^2\left(\left(r\partial_rg\right)^2+\left(\partial_\theta g\right)^2\right)+3l^3g^2\sin^2\theta\left(\left(r\partial_r\omega\right)^2+\left(\partial_\theta\omega\right)^2\right)
\nonumber \\
&-4f^2l^2g^2\Omega^2_N\left(\left(r\partial_r\psi\right)^2+\left(\partial_\theta\psi\right)^2\right)-4f^2l^2g^2\delta\left(\left(r\partial_r\chi\right)^2+\left(\partial_\theta\chi\right)^2\right)
\nonumber \\
&+4f^2lg^2\left(r\partial_rl+\cot\theta\partial_\theta l\right)-2f^2l^2gr\partial_rg+3l^3g^2\omega\sin^2\theta\left(\omega-2r\partial_r\omega\right)
\Biggr].
\end{align}

\begin{align}
\label{efe4}
r^2\partial^2_r\omega+\partial^2_\theta\omega &= \frac{1}{2}\frac{1}{fl}\Biggl[8flgm\Omega^2_N\psi^2\csc^2\theta\left(\omega_sr+m\omega\right)+4l\left(r^2\partial_rf\partial_r\omega+\partial_\theta f\partial_\theta\omega\right)
\nonumber \\
&-4l\omega r\partial_rf+f\left(4l+3r\partial_rl\right)\left(\omega-r\partial_r\omega\right)-3f\partial_\theta\omega\left(\partial_\theta l+2l\cot\theta\right)
\Biggr].
\end{align}

\begin{align}
\label{eompsi}
r^2\partial^2_r\psi+\partial^2_\theta\psi &= \frac{1}{2}\frac{1}{f^2l\Omega^2_N}\Biggl[2fl^2gr^2\partial_\psi V
-lg\psi\left((\omega_sr+m\omega)^2l-m^2f^2\csc^2\theta\right)\left(\psi\partial_\psi\Omega^2_N+2\Omega^2_N\right)
\nonumber \\
&-f^2l\partial_\psi\Omega^2_N\left(\left(r\partial_r\psi\right)^2+\left(\partial_\theta\psi\right)^2\right)-2f^2l\partial_\chi\Omega^2_N\left(r^2\partial_r\psi\partial_r\chi+\partial_\theta\psi\partial_\theta\chi\right)
\nonumber \\
&-f^2\Omega^2_Nr\partial_r\psi\left(r\partial_rl+4l\right)-f^2\Omega^2_N\partial_\theta\psi\left(\partial_\theta l +2l\cot\theta\right)
\Biggr].
\end{align}

\begin{align}
\label{eomchi}
\delta\left(r^2\partial^2_r\chi+\partial^2_\theta\chi\right) &= \frac{1}{2}\frac{1}{f^2l}\Biggl[2fl^2gr^2\partial_\chi V-lg\psi^2\partial_\chi\Omega^2_N\left((\omega_sr+m\omega)^2l-m^2f^2\csc^2\theta\right)+f^2l\partial_\chi\Omega^2_N\left(\left(r\partial_r\psi\right)^2+\left(\partial_\theta\psi\right)^2\right)
\nonumber \\
&-f^2r\partial_r\chi\delta\left(r\partial_rl+4l\right)-f^2\partial_\theta\chi\delta\left(\partial_\theta l+2l\cot\theta\right)
\Biggr].
\end{align}

\bibliographystyle{ieeetr}
\bibliography{biblio}

\end{document}